 \documentclass[final,3p,times]{elsarticle}
\usepackage{amsmath,amsfonts,amssymb,enumerate,multicol,}
\usepackage{tikz}
\usetikzlibrary{quantikz}
\usepackage{multirow}
\usepackage{graphicx,rotating,booktabs}
\usepackage{siunitx}
\usepackage{tabularx}

\usepackage{blindtext}
\usepackage{natbib}
\biboptions{sort&compress}
\newcolumntype{L}[1]{>{\raggedright\let\newline\\\arraybackslash\hspace{0pt}}m{#1}}
\newcolumntype{C}[1]{>{\centering\let\newline\\\arraybackslash\hspace{0pt}}m{#1}}
\newcolumntype{R}[1]{>{\raggedleft\let\newline\\\arraybackslash\hspace{0pt}}m{#1}}
\makeatletter

\renewcommand{\fnum@figure}{Fig. \thefigure}
\makeatother
\journal{}

\begin{document}

\begin{frontmatter}

\title{Quantum secret sharing using GHZ state qubit positioning and selective qubits strategy for secret reconstruction}
\author[]{Farhan Musanna}
\ead{fmusanna@ma.iitr.ac.in}
\author[]{Sanjeev Kumar\corref{cor1}}
\ead{malikfma@iitr.ac.in}
\cortext[cor1]{Corresponding author}
\address{Department of Mathematics \\
    Indian Institute of Technology Roorkee, Roorkee-247667, India \\
    Tel.: +91-1332-285853\\
    Fax: +123-45-678910\\}

\begin{abstract}
The work presents a novel quantum secret sharing strategy based on GHZ product state sharing between three parties. The dealer, based on the classical information to be shared, toggles his qubit and shares the product state. The other parties make their Bell measurements and collude to reconstruct the secret. Unlike the other protocols, this protocol does not involve the entire initial state reconstruction, rather uses selective qubits to discard the redundant qubits at the time of reconstruction to decrypt the secret. The protocol also allows for security against malicious attacks by an adversary without affecting the integrity of the secret. The security of the protocol lies in the fact that each party's correct announcement of their measurement is required for reconstruction, failing which the reconstruction process is jeopardized, thereby ascertaining the $(3,3)$ scheme which can further be extended for a $(n,n)$ scheme.
\end{abstract}
\end{frontmatter}
\section{Introduction}
As an amalgam of classical cryptography and quantum mechanics, quantum cryptography has the potential to achieve perfect security. The striking part about the security of this paradigm lies not in the complex mathematical equations involved in classical cryptography, but in the quantum mechanical laws that govern them. As an improvisation to traditional cryptography, secret sharing emerged as an efficient tool for secure data distribution and secret reconstruction. Traditional algorithms by the academia were given based on the concept given in the seminal paper \cite{shamir1979share} by Adi Shamir 'How to share a secret' in 1979. The foundations of the quantum counterpart of the classical version were first put by Hillery et al. in \cite{hillery1999quantum}. The work forms the basis of most quantum secret sharing algorithm currently devised. The work was further studied for the entanglement properties and measurement basis directions to be incorporated in the works by Karlsson et al. in \cite{karlsson1999quantum}. The scheme went one step further to explain the concept of eavesdropping detection and mitigation. The scheme proposed by Xiao et al. in \cite{xiao2004efficient} was a generalization of Hillery's scheme into arbitrary multi parties. Zhang et al. in \cite{zhang2005multiparty} used a single particle for multiparty secret sharing using different single-qubit gates and $x$ and $z$ direction basis. Guo \& Guo, in their work presented in \cite{guo2003quantum}, implemented quantum secret sharing without entanglement and claimed $100\%$ theoretical security. Focussing on more practical aspects, Tittel et al. in \cite{tittel2001experimental} demonstrated the concept of quantum secret sharing experimentally. Zhang et al. in \cite{zhang2005multiparty} and Li et al. in \cite{li2004multiparty} harnessed the feature of qubit swapping and product states formed by Bell state to devise a novel secret sharing algorithm. Rong et al. gave a novel idea in \cite{zu2011quantum} that proposed secret sharing with the aid of quantum error-correcting codes. To prevent any noise interruption during the communication of particles during secret sharing, many fault-tolerant schemes were proposed as in \cite{gu2010fault,yang2011fault} that were based on Decoherence-free subspace. There is another paradigm that deals with classical and quantum secret sharing working in tandem. These schemes rely on three phases; the first phase relies on using classical cryptographic techniques like Lagrange's interpolation over finite fields, Chinese remainder theorem on the secret for creating the shares. The second phase involves implementing the generalized Pauli operators on the individual qubits for particle distribution and doing a Quantum Fourier Transform (QFT) on individual qubits. The algorithms implementing this techniques include \cite{qin2016verifiable,qin2018multi,xiao2013multi,yang2013secret,mashhadi2019general}. In recent times, experimental work has picked up momentum, and works like \cite{williams2019quantum,zhou2018quantum} have been reported that involved polarization-entangled state and multipartite bound entanglement.
    \par The proposed work is based on the quantum mechanical properties of entanglement and the measurement of particles in the right Bell basis. Four fundamental single qubit gates are used for the transmission of the classical secret : $\mathbb{I}=\ket{0}\bra{0}+\ket{1}\bra{1}$, that does nothing on the quantum state, $\mathbb{X}= \ket{0}\bra{1}+\ket{1}\bra{0}$, that acts as a NOT gate and flips the qubit that is acted upon, $\mathbb{Y}= i\ket{1}\bra{0}-i\ket{0}\bra{1}$ acts as rotation about the $Y$ axis by $\pi$ on the Bloch sphere, $\mathbb{Z}= \ket{0}\bra{0}-\ket{1}\bra{1}$ which acts as a rotation about the $Z$ axis by $\pi$.
\section{Protocol}
The protocol consists of $4$ participants: the dealer $D$ and three reconstructors $P_1,P_2,P_3$. The dealer generates and shares at random a product state from the following: \begin{equation}\label{state}
\begin{aligned}
  \ket{A}= {} &\left(\dfrac{\ket{000}_{123}+\ket{111}_{123}}{\sqrt{2}}\right)\left(\dfrac{\ket{000}_{456}+\ket{111}_{456}}{\sqrt{2}}\right)\\
  \ket{B}=  &\left(\dfrac{\ket{001}_{123}+\ket{110}_{123}}{\sqrt{2}}\right)\left(\dfrac{\ket{001}_{456}+\ket{110}_{456}}{\sqrt{2}}\right)\\
    \ket{C}= &  \left(\dfrac{\ket{011}_{123}+\ket{100}_{123}}{\sqrt{2}}\right)\left(\dfrac{\ket{011}_{456}+\ket{100}_{456}}{\sqrt{2}}\right)\\
        \ket{D}=  & \left(\dfrac{\ket{101}_{123}+\ket{010}_{123}}{\sqrt{2}}\right)\left(\dfrac{\ket{101}_{456}+\ket{010}_{456}}{\sqrt{2}}\right)
          \end{aligned}
          \end{equation} The indexing of the qubits denote the particles in possession of players, i.e $P_1$ has $1$ and $6$, $P_2$ has $2$ and $5$, $P_3$ has $3$ and $4$. $D$ chooses to operate single qubit  gates $\mathbb{I},\mathbb{X},i\mathbb{Y},\mathbb{Z}$ on the qubits to share the classical secret in the following manner : \begin{eqnarray*}
          \mathbb{I}_1 &=& `00', \mathbb{I}_6 = `11'  \\
          \mathbb{X}_1 &=& `01', \mathbb{X}_6 = `10' \\
           i\mathbb{Y}_1 &=& `11', i\mathbb{Y}_6 = `00'\\
            \mathbb{Z}_1&=& `10', \mathbb{Z}_6 = `01'
            \end{eqnarray*} for ex: $\mathbb{Z}_1$ and $\mathbb{Z}_6$ denotes a $\mathbb{Z}$ gate on qubit $1$ and $6$ respectively. The secret sharing involves the following steps:
            \begin{enumerate}
            \item $D$ randomly chooses to operate on qubit $1^{st}$ or $6^{th}$ and reserves this announcement until the other parties ask him.
            \item $P_1$ makes Bell basis measurement on his qubits pair $(1,6)$ and reserves his measurement results till further notice.
            \item $P_2$ and $P_3$ also make respective Bell measurements on their qubits and do not disclose their results.
            \item The reconstruction involves each part to give in their measurement results starting with $D$'s information about the state created and the qubit toggled.
            \end{enumerate}
           \section{Example}
           \begin{enumerate}
           \item[Step 1.] Suppose $D$ creates the product state out of GHZ states \cite{greenberger1989bell}, \begin{equation*}\ket{A}=\left(\dfrac{\ket{000}_{123}+\ket{111}_{123}}{\sqrt{2}}\right)\left(\dfrac{\ket{000}_{456}+\ket{111}_{456}}{\sqrt{2}}\right)
\end{equation*}
\item[Step 2.] $D$ operates a $i\mathbb{Y}$ gate on qubit $1$ and shares the qubits to the three participants.
\item[Step 3.] $P_1$ measures say $\ket{\beta^+}_{16}$, this collapses the system into the states \begin{equation}\label{P1}
\left(-\ket{\alpha^+}_{23}\ket{\alpha^-}_{45}- \ket{\alpha^-}_{23}\ket{\alpha^+}_{45}\right)
\end{equation} The set of all possible measurements outcomes are given in Table below

\begin{table}[!ht]
\centering
\begin{tabular}{|L{1.5cm}|L{1.5cm}|c|}
\hline
D's Operation $\rightarrow$ & $P_1$'s Outcome & Collapsed State  \\ \hline
\multirow{ 4}{*}{$\mathbb{I}_1$} & $\ket{\alpha^+}_{16}$ & $\dfrac{1}{2}\left(\ket{\alpha^+}_{23}\ket{\alpha^+}_{45}+ \ket{\alpha^-}_{23}\ket{\alpha^-}_{45}\right)$ \\
& $\ket{\alpha^-}_{16}$ & $\dfrac{1}{2}\left(\ket{\alpha^+}_{23}\ket{\alpha^-}_{45}+ \ket{\alpha^-}_{23}\ket{\alpha^+}_{45}\right)$\\
& $\ket{\beta^+}_{16}$ & $\dfrac{1}{2}\left(\ket{\beta^+}_{23}\ket{\beta^+}_{45}+ \ket{\beta^-}_{23}\ket{\beta^-}_{45}\right)$\\
& $\ket{\beta^-}_{16}$ & $\dfrac{1}{2}\left(\ket{\beta^+}_{23}\ket{\beta^-}_{45}+ \ket{\beta^-}_{23}\ket{\beta^+}_{45}\right)$ \\ \hline

\multirow{ 4}{*}{$\mathbb{X}_1$} & $\ket{\alpha^+}_{16}$& $\dfrac{1}{2}\left(\ket{\beta^+}_{23}\ket{\beta^+}_{45}+ \ket{\beta^-}_{23}\ket{\beta^-}_{45}\right)$ \\
& $\ket{\alpha^-}_{16}$ & $\dfrac{1}{2}\left(-\ket{\beta^+}_{23}\ket{\beta^-}_{45}- \ket{\beta^-}_{23}\ket{\beta^+}_{45}\right)$\\
& $\ket{\beta^+}_{16}$ & $\dfrac{1}{2}\left(\ket{\alpha^+}_{23}\ket{\alpha^+}_{45}+ \ket{\alpha^-}_{23}\ket{\alpha^-}_{45}\right)$ \\
& $\ket{\beta^-}_{16}$ & $\dfrac{1}{2}\left(\ket{-\alpha^+}_{23}\ket{\alpha^-}_{45}- \ket{\alpha^-}_{23}\ket{\alpha^+}_{45}\right)$ \\\hline

\multirow{ 4}{*}{$i\mathbb{Y}_1$} & $\ket{\alpha^+}_{16}$ & $\dfrac{1}{2}\left(-\ket{\beta^+}_{23}\ket{\beta^-}_{45}- \ket{\beta^-}_{23}\ket{\beta^+}_{45}\right)$ \\
& $\ket{\alpha^-}_{16}$ & $\dfrac{1}{2}\left(\ket{\beta^+}_{23}\ket{\beta^+}_{45}+ \ket{\beta^-}_{23}\ket{\beta^-}_{45}\right)$  \\
& $\ket{\beta^+}_{16}$ & $\dfrac{1}{2}\left(-\ket{\alpha^+}_{23}\ket{\alpha^-}_{45}- \ket{\alpha^-}_{23}\ket{\alpha^+}_{45}\right)$\\
& $\ket{\beta^-}_{16}$ & $\dfrac{1}{2}\left(\ket{\alpha^+}_{23}\ket{\alpha^+}_{45}+ \ket{\alpha^-}_{23}\ket{\alpha^-}_{45}\right)$  \\\hline

\multirow{ 4}{*}{$\mathbb{Z}_1$} &  $\ket{\alpha^+}_{16}$ &$\dfrac{1}{2}\left(\ket{\alpha^+}_{23}\ket{\alpha^-}_{45}+ \ket{\alpha^-}_{25}\ket{\alpha^+}_{45}\right)$\\
& $\ket{\alpha^-}_{16}$ & $\dfrac{1}{2}\left(\ket{\alpha^+}_{25}\ket{\alpha^+}_{45}+ \ket{\alpha^-}_{25}\ket{\alpha^-}_{45}\right)$ \\
& $\ket{\beta^+}_{16}$ & $\dfrac{1}{2}\left(\ket{\beta^+}_{25}\ket{\beta^-}_{45}+ \ket{\beta^-}_{25}\ket{\beta^+}_{45}\right)$ \\
& $\ket{\beta^-}_{16}$ & $\dfrac{1}{2}\left(\ket{\beta^+}_{25}\ket{\beta^+}_{45}+ \ket{\beta^-}_{25}\ket{\beta^-}_{45}\right)$\\
\hline
\end{tabular}
\caption{Unitary operation and Corresponding Measurement Results}
\label{Summary}
\end{table}
\item[Step 4.] For the reconstruction of the secret, $P_2$ and $P_3$ divulge in their measurement results say $\ket{\alpha^-}_{25}$ and $\ket{\alpha^+}_{34}$. The state produced by their measurements( without the normalizing factor) is \begin{equation}\label{ff}
    \begin{aligned}
    \ket{a}={} &\left( \ket{0}_2\ket{0}_5-\ket{1}_2\ket{1}_5 \right)\left( \ket{0}_3\ket{0}_4+\ket{1}_3\ket{1}_4 \right)\\
           = &\ket{0}_2\ket{0}_3\ket{0}_4\ket{0}_5+  \ket{0}_2\ket{1}_3\ket{1}_4\ket{0}_5- \ket{1}_2\ket{0}_3\ket{0}_4\ket{1}_5- \ket{1}_2\ket{1}_3\ket{1}_4\ket{1}_5
     \end{aligned}
     \end{equation}
  \item[Step 5.] Participants $P_2$ and $P_3$ ask $D$ for the state initially prepared. $D$ tells that it created the first or second state from eq.\eqref{state}, they keep the first and fourth term of the above equation else they keep the second and third term. Say $D$ created the first state so the state left after discarding the terms is :
      \begin{equation}
       \ket{0}_2\ket{0}_3\ket{0}_4\ket{0}_5- \ket{1}_2\ket{1}_3\ket{1}_4\ket{1}_5
\end{equation}
\item[Step 6.] $P_1$ divulges its share say $\ket{\beta^+}_{16}$ to get the state as:\begin{equation}\label{fr}
\begin{aligned}
 \ket{b}={} & (\ket{0}_1\ket{1}_6+\ket{1}_1\ket{0}_6)(\ket{0}_2\ket{0}_3\ket{0}_4\ket{0}_5- \ket{1}_2\ket{1}_3\ket{1}_4\ket{1}_5)\\
 =& \ket{0}_1\ket{0}_2\ket{0}_3\ket{0}_4\ket{0}_5\ket{1}_6- \ket{0}_1\ket{1}_2\ket{1}_3\ket{1}_4\ket{1}_5\ket{1}_6+ \ket{1}_1\ket{0}_2\ket{0}_3\ket{0}_4\ket{0}_5\ket{0}_6 - \ket{1}_1\ket{1}_2\ket{1}_3\ket{1}_4\ket{1}_5\ket{0}_6
 \end{aligned}
                                                                             \end{equation}
 \item[Step 7.] $D$ confirms which state it generated and which qubit it changed. If $D$ says for instance it created the first state from eq.\eqref{state} and it toggled the first qubit, then they know that qubit $4,5,6$ should be in state $\ket{0}\ket{0}\ket{0}$ or $\ket{1}\ket{1}\ket{1}$, so they discard the first and fourth term from the above eq.\eqref{fr} and get \begin{equation}\label{dd}
                                                        \ket{c}= \ket{1}_1\ket{0}_2\ket{0}_3\ket{0}_4\ket{0}_5\ket{0}_6 - \ket{0}_1\ket{1}_2\ket{1}_3\ket{1}_4\ket{1}_5\ket{1}_6
                                                      \end{equation}
                                                      \item[Step 8.] Inferring the fact that the terms represent the two terms with perfect correlation between the qubits of the first product state from eq. \eqref{state}, they conclude that the operation done is $i\mathbb{Y}$, since
     \begin{equation}\label{iy}
     i\mathbb{Y}_1(\ket{0}_1\ket{0}_2\ket{0}_3\ket{0}_4\ket{0}_5\ket{0}_6 + \ket{1}_1\ket{1}_2\ket{1}_3\ket{1}_4\ket{1}_5\ket{1}_6)= \ket{c}= \ket{1}_1\ket{0}_2\ket{0}_3\ket{0}_4\ket{0}_5\ket{0}_6 - \ket{0}_1\ket{1}_2\ket{1}_3\ket{1}_4\ket{1}_5\ket{1}_6
     \end{equation}
     \item[Step 9.] Thus, the parties, by the information ushered by the participants and discarding strategy of qubits reconstruct the secret operator and hence the secret message i.e `11', without reconstructing the initial state.
         \end{enumerate}
         \section{Security Analysis}
         The security analysis of the protocol is based on the information that is not made public by $D$ and other participants.
         \subsection{$D$ hides the state being shared}
         Suppose $D$ created the state $\ket{C}$ and applied an $\mathbb{X}_1$ gate, then $P_1$'s measurement of say $\ket{\alpha^+}_{16}$ will collapse the system into the state $\dfrac{1}{2}\left(\ket{\alpha^+}_{23}\ket{\alpha^+}_{45}- \ket{\alpha^-}_{23}\ket{\alpha^-}_{45}\right)$,  which can be seen in a row corresponding to every operator (apart from the phase of the state) in Table \ref{Summary}. If $D$ announces incorrectly that he created the state $\ket{A}$, then parties collude to deduce the secret as $\mathbb{I}_1$ instead of the real secret $\mathbb{X}_1$, i.e. `01'. The same is the case when the state is $\ket{B}$ and $\ket{D}$. So in this scenario, the secret will never be revealed, and hence the participation of $D$ is necessary.
         \subsection{D lies about the qubit acted upon}
           $D$ has the discretion to choose between the qubits. Suppose $D$ announces correctly the state it prepared but lies about the qubit. For instance if $D$ toggled the $1^{st}$ qubit, but announces the $6^{th}$ qubit, then according to Step 6 of the preceding example, the parties know that qubits $1,2,3$ would be in states $\ket{0}\ket{0}\ket{0}$ or $\ket{1}\ket{1}\ket{1}$ and hence they agree on state \begin{equation}\label{dd}
                                                                                               \ket{b}=\ket{0}_1\ket{0}_2\ket{0}_3\ket{0}_4\ket{0}_5\ket{1}_6- \ket{1}_1\ket{1}_2\ket{1}_3\ket{1}_4\ket{1}_5\ket{0}_6
                                                                                             \end{equation}. Deciphering on the same grounds of Step 8, they conclude that the operation was actual $i\mathbb{Y}_6$. Though they guess the operator correctly, the qubit operated upon makes them their final conclusion of the classical secret `00' wrong. So, the protocol is robust in the sense that the position of the qubit acted upon and $D$'s correct announcement is fundamental to the correct message to b communicated.
           \subsection{$P_1$ hides about his measurement results}
             The announcement of $P$'s measurement outcome is crucial in the reconstruction of the secret. This is ascertained by the following two cases (i) $P_1$ does not give information about the measurement, the consequences are obvious from  Table \ref{Summary}. For instance, every collapsed state in the third column corresponds to every measurement outcome of $P_1$ and therefore to every operator either $\mathbb{I}_1, \mathbb{X}_1 ,i\mathbb{Y}_1 ,\mathbb{Z}_1$, for instance  \begin{equation*}
             \begin{aligned}
             \mathbb{I}_1 \Leftrightarrow    {}& \ket{\alpha^-}_{16}\left(\ket{\alpha^+}_{23}\ket{\alpha^-}_{45}+ \ket{\alpha^-}_{23}\ket{\alpha^+}_{45}\right)\\
             \mathbb{X}_1\Leftrightarrow &  \ket{\beta^-}_{16}\left(\ket{\alpha^+}_{23}\ket{\alpha^-}_{45}+ \ket{\alpha^-}_{23}\ket{\alpha^+}_{45}\right)\\
             i\mathbb{Y}_1\Leftrightarrow & \ket{\beta^+}_{16}\left(\ket{\alpha^+}_{23}\ket{\alpha^-}_{45}+ \ket{\alpha^-}_{23}\ket{\alpha^+}_{45}\right)\\
             \mathbb{Z}_1\Leftrightarrow & \ket{\alpha^+}_{16}\left(\ket{\alpha^+}_{23}\ket{\alpha^-}_{45}+\ket{\alpha^-}_{23}\ket{\alpha^+}_{45}\right)
             \end{aligned}
             \end{equation*}
             \subsection{$P_3$ does not collude}
             Collusion of $P_2$ and $P_3$ is necessary for the reconstruction of the fact. Consider Step $2$ of the previous section. The simplified form of eq.\eqref{P1} is \begin{equation}\label{aa}
                                \ket{\chi}=\left(\ket{\alpha^+}_{25}\ket{\alpha^-}_{34}+\ket{\alpha^-}_{25}\ket{\alpha^+}_{34}\right)
                              \end{equation}If $D$ had operated with the $\mathbb{I}_1$ gate, then the simplified equation would have been  \begin{equation}\label{aaa}
                                \ket{\chi}=\left(\ket{\alpha^+}_{25}\ket{\alpha^+}_{34}+\ket{\alpha^+}_{25}\ket{\alpha^+}_{34}\right)
                              \end{equation} If $P_3$ refrains from sharing his announcement result, then for either measurement result $\ket{\alpha^+}_{25}$ or $\ket{\alpha^-}_{25}$ $P_3$ will be in a dilemma whether $P_3$ got a $\ket{\alpha^+}_{34}$ or $\ket{\alpha^-}_{34}$ which would lead him to incorrectly decrypt the operator as $i\mathbb{Y}_1$ or $\mathbb{X}$. Hence, alone $P_2$ cannot reconstruct the secret.
                \subsection{Adversary Eve's attack on the state}
                 An adversary may intend to disrupt the secret reconstruction by altering the state prepared. For instance if $D$ prepares the state $\ket{A}$, operates a $\mathbb{Z}_1$ gate and shares the state. Eve intercepts the transmission and modifies the $6^{th}$ qubit with an arbitrary gate say $\mathbb{X}_6$, so the modified state looks like
                 \begin{equation}\label{aaa}
                   \ket{\tilde{A}}=\left(\dfrac{\ket{000}_{123}-\ket{111}_{123}}{\sqrt{2}}\right)\left(\dfrac{\ket{001}_{456}+\ket{110}_{456}}{\sqrt{2}}\right)
                 \end{equation} $P_1$ measures his particles for the state say $\ket{\alpha^+}_{16}$ which collapses the system into
                 \begin{equation*}
                  \ket{\beta^+}_{25}\ket{\beta^-}_{34}+\ket{\beta^-}_{25}\ket{\beta^+}_{34}
                  \end{equation*}  $P_2$ and $P_3$ measure $\ket{\beta^-}_{25}$ and $\ket{\beta^+}_{34}$ to obtain
                  \begin{equation*}
                  \ket{0}_2\ket{0}_3\ket{1}_4\ket{1}_5 + \ket{0}_2\ket{1}_3\ket{0}_4\ket{1}_5 - \ket{1}_2\ket{0}_3\ket{1}_4\ket{0}_5- \ket{1}_2\ket{1}_3\ket{0}_4\ket{0}_5
                  \end{equation*} They implement the discarding strategy to keep the first and fourth term \begin{equation*}
                                                                                                             \ket{0}_2\ket{0}_3\ket{1}_4\ket{1}_5- \ket{1}_2\ket{1}_3\ket{0}_4\ket{0}_5
                                                                                                           \end{equation*} They ask for $P_1$'s measurement of $\ket{\alpha^+}_{16}$ to get
                  \begin{equation*}
                   \ket{0}_1\ket{0}_2\ket{0}_3\ket{1}_4\ket{1}_5\ket{0}_6- \ket{0}_1\ket{1}_2\ket{1}_3\ket{0}_4\ket{0}_5\ket{0}_6+ \ket{1}_1\ket{0}_2\ket{0}_3\ket{1}_4\ket{1}_5\ket{1}_6- \ket{1}_1\ket{1}_2\ket{1}_3\ket{0}_4\ket{0}_5\ket{1}_6
                   \end{equation*}
                    Now, they ask $D$ about the state he prepared and the qubit he transformed, which he obliges to. Using this knowledge of the positioning of qubits and the logic used in Step 7, they keep the second and third term from the above equation and deduce the operation $i\mathbb{Y}_1$. Since they have the state with them, $P_2$ and $P_3$ also get to know from the discarded terms, the operator Eve used on its qubit i.e $\mathbb{X}_1$. One important aspect of reconstructing the secret is the correct announcement of $D$ of the qubit it transformed. If $D$ said it transformed the $6^{th}$ qubit, then $P_2$ and $P_3$ would have kept the first and fourth terms to deduce the operator as $\mathbb{X}_6$.
                    \section{Conclusion}
                         The proposed protocol utilizes the entangled GHZ state for secret sharing between three parties and the dealer. The participation of each party with his share is mandatory for reconstruction, as has been shown in the security analysis. Any eves-dropping on a particular state is identified and is discarded, to arrive at the real secret. The judicious discarding strategy by the players eliminates the spurious cases and facilitates in the secret reconstruction without reconstructing the entire original state. The other key feature of the algorithm lies in the communication of the classical message based on the qubit positioning, which adds an extra level of security since the same operator encodes two different messages based on the qubit it acted upon. The current scheme has scope for the generalized $(n,n)$ secret sharing techniques for transmitting classical information in areas of quantum image processing and many more.

%\bibliographystyle{unsrt}
%\bibliography{ipl}

\end{document}